\documentstyle[epsfig,12pt]{article}
\begin{document}

\newcommand \be  {\begin{equation}}
\newcommand \bea {\begin{eqnarray} \nonumber }
\newcommand \ee  {\end{equation}}
\newcommand \eea {\end{eqnarray}}

\title{{\bf Anomalous dynamical light scattering in soft glassy gels}}

\author{J.-P. Bouchaud$^{1}$, E. Pitard$^2$}

\date{\it
$^1$ Service de Physique de l'\'Etat Condens\'e,\\
 Centre d'\'etudes de Saclay, \\ Orme des Merisiers, 
91191 Gif-sur-Yvette Cedex, France
\\
$^2$ Laboratoire
de Physique Math\'ematique et Th\'eorique,\\ 
Universit\'e Montpellier II, UMR 5825, France  
}
\maketitle

\begin{abstract}
We compute the dynamical structure factor $S(q,\tau)$ 
of an elastic medium where
force dipoles appear at random in space and in time, due to `micro-collapses' of the 
structure.
Various regimes are found, depending on the
wave vector $q$ and the collapse time $\theta$. In an early time regime, the 
logarithm
of the structure factor behaves as $(q\tau)^{3/2}$, as predicted in 
\cite{Cip} using heuristic
arguments. However, in an intermediate time regime we rather obtain a $(q 
\tau)^{5/4}$ behaviour.
Finally, the asymptotic long time regime is found to behave
as $q^{3/2} \tau$. We also give a plausible scenario for aging, in terms of a 
strain
dependent energy barrier for micro-collapses. The relaxation time is found to 
grow with
the age $t_w$, quasi-exponentially at first, and then as $t_w^{4/5}$ with 
logarithmic corrections.
\end{abstract}

\vskip 1cm

\section{Introduction}

There is currently a great interest in soft glassy materials, such as pastes,
foams or colloidal gels \cite{Sollich}. 
These systems are often out-of-equilibrium and exhibit 
interesting aging effects, typical of other glassy systems such as spin-glasses
or polymer glasses \cite{Revus,StAndrews}. For example, their rheological 
properties
are found to depend strongly on the age of the system \cite{Derec,Cloitre}. 
From a theoretical point of view, aging is expected both in response functions
(such as magnetic susceptibilities or elastic moduli) but also in correlation
functions, which are notoriously harder to study experimentally. Recently, a 
new experimental `multi-speckle' technique
has allowed one to properly investigate aging in a dynamical correlation 
function
\cite{Cip}.
The difficulty is to make the experimental acquisition time much smaller than 
the
typical age of the system, otherwise the age of the system changes 
significantly
during the experiment and the measured correlation function is meaningless. 
Using this technique, the dynamical structure factor of a colloidal gel made 
of
aggregating polystyrene particles was measured
and revealed several unexpected features \cite{Cip}. (i) For a given waiting 
time $t_w$
(counted from the moment when the gel is formed), the dynamical structure 
factor
$S(q,\tau,t_w)$ is found to decay as $\exp(-A(q \tau)^{3/2})$. This must be 
contrasted
with the usual diffusive decay, as $\exp(-Dq^2 \tau)$: both the $q$ and $\tau$ 
dependence are anomalous, and non intuitive. The exponent $3/2$ shows that the 
time decay
is {\it faster} than an exponential, whereas one would have expected a slow 
decay in a
glassy system. A similar exponent $3/2$ was also found in different systems,
such as diblock polystyrene/polyisoprene copolymers \cite{Lairez}, laponite 
\cite{Lequeux}
or other systems \cite{Cipelse}. (ii) The scaling in 
$q\tau$ suggests that some kind of convection, rather 
than diffusion, is present in the system. However, since the system is density 
matched,
this cannot be a global sedimentation effect. (iii) When the age of the system 
increases,
the relaxation of the dynamical structure factor becomes slower but retains 
its shape.
In other words, only the coefficient $A$ in the above expression is found to 
be age
dependent. The corresponding relaxation time $\tau_r(q,t_w)$ (defined as 
$A(q\tau_r)^{3/2}=1$) is found to increase {\it exponentially} with $t_w$ at 
first,
and then as $t_w^\mu$ with $\mu < 1$ for larger waiting times. Such an 
exponentially
growing relaxation time with age was also reported for laponite \cite{Abou}, 
but is
very unusual. The second regime, however, is typical of most experimental 
glassy
systems \cite{Struik,Sitges}.

A heuristic interpretation of the $(q \tau)^{3/2}$ scaling was proposed in 
\cite{Cip}.
The basic mechanism is that the gel contracts in a very heterogeneous way: 
localized
`micro-collapses' create a long range elastic deformation field which is 
ultimately
responsible for the $q^{3/2}$ behaviour. In order to get the correct $\tau$ 
dependence
the strain within these collapsing regions was postulated to be linear in 
$\tau$. This
is however awkward since the instant $\tau=0$ has no special meaning and a 
linear growth
would eventually lead to unbounded strains. Furthermore, this argument does 
not
account for aging.

The aim of this paper is to put the arguments of \cite{Cip} on a firmer 
footing. We
show that an important time scale was left out in the analysis of \cite{Cip}, 
namely
the typical collapse time which we will call $\theta$. We find that for 
$\tau \ll \theta$, the relaxation of the structure factor for a given $t_w$ 
does
indeed only depend on $q \tau$, but exhibits two distinct regimes: a short time
regime where we recover the $(q \tau)^{3/2}$ behaviour, and an intermediate 
time
regime where the decay is slower, as $\exp(-B(q \tau)^{5/4})$, which might be 
of experimental
relevance. 
For very large $\tau$'s, the $q \tau$ scaling breaks down and 
a new dynamical regime is found.  
We also give a simple argument to rationalize the dependence of $\tau_r$ with 
$t_w$.

\section{The model}

Following \cite{Cip}, we assume that the dominant mechanism is the random 
appearance
of micro-collapses: since the micro-particles forming the gel attract each 
other rather
strongly, the gel tends to restructure locally so as to create dense packing 
of particles.
(Post mortem analysis indeed reveals that particles actually tend to {\it 
fuse} together).
Since the collapsing particles belong to a gel network, their motion will 
induce a certain strain
field around them; other particles therefore move and dynamical light 
scattering probes this
motion. Assume for a moment that the gel is a one dimensional chain of beads 
and springs.
When two particles decide to leave their equilibrium position to stick 
together, the left
particle imposes a positive force $+f_0$ on the left part of the chain, while 
the
right particle
imposes an equal opposite force $-f_0$ on the right part of the chain. In 
other words, this
creates a {\it dipole} of forces. This is also true in three dimensions, where 
the collapse
of particles will result in a force dipole of intensity $P_0$ in the direction 
$\vec n$. When the
dipole is formed at point $\vec r_0$, the elastic strain field $\vec u$ at 
point $\vec r$ can
be computed assuming a simple central force elasticity:
\be
K \Delta \vec u = - \vec f (\vec r) ,\label{eqn1}
\ee
where $K$ is a compression modulus
and $\vec f (\vec r)$ is a local force dipole of the
form:
\be
\vec f(\vec r)=
f_0 \left[\vec \epsilon \cdot \vec \nabla \delta(\vec r - \vec r_0)\right] 
\vec n,
\ee
where we will assume that $\vec \epsilon = \epsilon \vec n$, i.e, the mean 
displacement
of the particles creates a force in the same direction. This dipolar
force can be simply expressed by its Fourier transform:
\be
\vec f_k= i P_0 \,(\vec k . \vec n)\, \vec n \, \exp(-i\vec k . \vec r_0)
\ee
with $P_0=f_0 \epsilon$.
Note that a more refined model with shear modulus could be
also be considered, but would only change some numerical factors in the 
following calculations.
The solution of equation (\ref{eqn1}) is of course:
\be
\vec u(\vec r) = - \frac{P_0}{4\pi K} \frac{(\vec r -\vec r_0)\cdot \vec 
n}{|r-r_0|^3} \vec n.
\ee
The $r^{-2}$ dependence of the strain field has an immediate consequence which 
will be
of importance in the following: if there is a finite density of force dipoles 
randomly
scattered in space, the probability that the stress at a given point has an 
amplitude $u$
decays for large $u$ as $u^{-5/2}$, which has a diverging variance. (This 
divergence is
however cut-off if the finite size of the dipoles is taken into account).
This property of the distribution of displacements and stresses will be
responsible for the unusual $q$-dependence of the structure factor. Note also
that from
dimensional arguments, $P_0/4\pi K=v_0$ is a volume given by $\xi^3 
\lambda^2$, where $\xi$
is the typical size of the collapsing region, and $\lambda < 1$ is the 
contraction ratio (that we will assume to be of order 1 in the following).
 
Now, let us assume that the micro-collapses are not instantaneous 
but take place progressively,
over a time scale $\theta$. 
A given event  $j$ starts at time $t_j$ and is completed at time $t_j+\theta$;
the dipole intensity $P(t)$ at time $t$ is a certain function of 
$(t-t_j)/\theta$.
We will assume 
that this function is approximately linear, and simply write $P_j(t)=P_0 
(t-t_j)/\theta$,
which saturates at
$P_0$ when the collapse is completed. (Note that a micro-collapse could itself 
be the
result of many successive events). The dynamics of the individual particles is 
presumably
dominated by viscous friction, therefore we write the following equation of 
motion for the
strain field $\vec u$:
\be
\gamma \frac{\partial \vec u(\vec r,t)}{\partial t} =  
K \Delta \vec u + \sum_j \vec f_{j}(\vec r, t) 
 + \vec \eta(\vec r,t),\label{eqmotion}
 \ee
 where the Fourier transform of the dipolar force is
\be
\vec f_{j}(\vec k, t)=i P_j(t) \,(\vec k . \vec n_j)\, \vec n_j \, \exp(-i\vec 
k. \vec r_j)
\ee
and $\gamma$ is a friction coefficient, $\vec \eta$ is the thermal random 
force,
 and we now take into account the fact that many 
microcollapses take place, at different times $t_j$
and different positions $\vec r_j$, with different orientations $\vec n_j$ of
the dipoles. 
Actually, we will assume in the following that these events occur randomly in 
space and time,
with a certain rate $\rho$ per unit volume and unit time (see below for a 
further discussion of this 
assumption). The quantity $K/\gamma$ is a diffusion constant that we will call 
$D$. Equation
(\ref{eqmotion}) defines the model that we want to study and from which we 
will compute
the dynamical structure factor $S(q,\tau)$, defined as:
\be
S(q,\tau) = 
\langle \exp\left[i \vec q \cdot \left(\vec u(\vec r,t+\tau) 
- \vec u(\vec r,t)\right)\right] \rangle
\ee
where the brackets refer to a spatial average over $\vec r$ or, equivalently, 
over the
random location and time of the micro-collapse events. In the following, we 
will neglect
the thermal random force, which would add a Debye-Waller diffusive 
contribution to the
dynamical structure factor, and set $\vec \eta = 0$. However, as we discuss 
below,
the presence of $\vec \eta$ has an indirect crucial effect since 
the nucleation of micro-collapses is probably of thermal origin.

\section{The slow collapse regime}

A first step is to calculate the Fourier transform of the {\it time 
derivative} of
the displacement field $\vec u(\vec r,t)$ created by a single dipole located 
at $\vec r_j$,
in direction $\vec n_j$, that we denote $\vec v(\vec k,t|\vec r_j,\vec 
n_j,t_j)$. One finds:
\bea
\vec v(\vec k,t|\vec r_j,t_j)& = & -i \exp(-i\vec k\cdot \vec r_j) 
\frac{P_0 \vec n_j}{\theta} \ \frac{\vec n_j \cdot \vec k}{K k^2}
\exp(-Dk^2 t)\\& & \left[\exp(Dk^2 t_j) - \exp(Dk^2 \min(t_j+\theta,t))\right].
\eea
The displacement difference between $t$ and $t+\tau$ can therefore be 
expressed as:
\be
\vec u(\vec r,t+\tau)-\vec(\vec r,t)=
\int_t^{t+\tau} dt' \sum_{i/t_j < t'} \int \frac{d^3 \vec k}{(2\pi)^3} 
\exp(i\vec k\cdot \vec r) \vec v(\vec k,t|\vec r_j,t_j)
\ee
The analysis of this expression reveals that there are {\it a priori} five 
different cases
to consider for the relative position of the time $t_j$ when the $j^{th}$ 
micro-collapse
takes place and the other relevant times: (a1) $t_j \leq t-\theta \leq t' 
-\theta$:
the $j^{th}$ event is over before $t,t'$, (a2) $t_j \leq t'-\theta$ 
and $t-\theta < t_j \leq t+\tau-\theta$; and (b1)  $t_j >  t'-\theta$, 
$t-\theta \leq t_j \leq t$
and $t_j < t+\tau-\theta$, (b2) $t_j >  t'-\theta$ and $t+\tau-\theta < t_j < 
t$ and
finally (b3) $t_j >  t'-\theta$ and $t< t_j < t+\tau$. 

We first consider the `slow' case
where the experimental time $\tau$ is very small compared to the collapse time 
$\theta$.
The contribution of the different regimes to $S(q,\tau)$ can be estimated and 
one finds
that a new, $q$-dependent time scale $\tau_q$ appears, defined as:
\be
\tau_q \equiv \frac{D\theta}{q v_0} \theta,
\ee
such that, depending on the ratio $\tau/\tau_q$, 
the dominant contribution to the decay of $S(q,\tau)$ comes from different 
regions of the $t',t_j$ plane. Let us first consider the case $\tau \ll 
\tau_q$.
The dominant contribution then comes from region (b2). Introducing 
$\tau_j=t-t_j$,
the average contribution of a event $i$ to $S(q,\tau)$ is given by:
\be
1 + 2\pi \int \frac{r^2 dr}{V} \int d(\cos \alpha) \int \frac{d\vec n}{4\pi} 
\left\{\exp\left[i(\vec q \cdot \vec n) \cos \alpha \frac{\tau v_0}{\theta r^2}
{\cal F}(\frac{r}{2\sqrt{D\tau_j}})\right] - 1\right\},\label{b2}
\ee
where $V$ is the total volume of the sample, and $\cal F$ is a function 
defined as:
\be
{\cal F}(x) = {\rm erfc}(x) - x \ {\rm erfc}'(x) \qquad {\cal F}(0) = 1.
\ee
Now, introducing  in Eq. (\ref{b2}) $r = s \sqrt{|\vec q \cdot \vec n|\tau 
v_0/\theta}$
and $\tau_j=z \theta$ with $0 \leq z \leq 1$, 
one can rewrite (\ref{b2}) for very large volumes $V$ as:
\be
\exp\left[\int d\vec n \left(\frac{|\vec q \cdot \vec n|\tau 
v_0}{\theta}\right)^{3/2}
\int \frac{s^2 ds}{V} \left\{\frac{s^2}{{\cal F}(\frac{s}{2}\sqrt{R})} 
\sin \frac{{\cal F}(\frac{s}{2}\sqrt{R})}{s^2} - 1 \right\} \right]
\ee 
with $R={\tau |\vec q \cdot \vec n|}/{\tau_q z}$. Each micro-collapse 
contributes
independently; and since within a small time interval $\theta dz$ one has a 
total of
$\rho V \theta dz$ events, the contribution of the (b2) regime reads:
\be 
\exp\left[\rho \theta \int_0^1 dz \int d\vec n 
\left(\frac{|\vec q \cdot \vec n|\tau v_0}{\theta}\right)^{3/2}
\int 
s^2 ds \left\{\frac{s^2}{{\cal F}(\frac{s}{2}\sqrt{R})} 
\sin \frac{{\cal F}(\frac{s}{2}\sqrt{R})}{s^2} - 1 \right\} \right].
\ee 
Since we have assumed that $\tau \ll \tau_q$, it is justified to set $R=0$ 
provided
all integrals (over $s$ and $z$) converge. One can check that this is indeed 
the case.
The final result reads (for $t > \theta$):
\be
S(q,\tau) = \exp\left[-\frac{16 \sqrt{2\pi^3}}{75} \rho \theta (D\theta)^{3/2}
\left(\frac{\tau}{\tau_q}\right)^{3/2}\right] \qquad (\tau \ll 
\tau_q),\label{result1}
\ee
which has the form suggested by the arguments of \cite{Cip}, in particular, it 
indeed only
depends on $(q\tau)^{3/2}$. (The numerical value of the prefactor is 
$1.67996..$.)
However, the dimensional factors differ from those 
obtained in \cite{Cip}, and the result is only valid in the short time regime 
$\tau \ll
\theta$ and $\tau \ll \tau_q$. Note that the combination $\hat \rho 
\equiv
\rho \theta (D\theta)^{3/2}$
is adimensional and represents the average number of events taking place 
within a time
interval $\theta$ and within a diffusion volume $(D\theta)^{3/2}$. The 
contribution of the
other regions in the $t',t_j$ plane can be analyzed similarly, and lead to 
sub-dominant
corrections proportional to $\hat \rho (\tau/\tau_q)^2$ and $\hat \rho 
(\tau/\tau_q)^{5/2}$.
Therefore, if $\tau/\tau_q$ is not very small, an effective exponent somewhat 
larger than
$3/2$ can be observed.

Turning to the regime $\tau_q \ll \tau \ll \theta$, we now find that the 
dominant regime is
(a1). Introducing $\tau_j=t-\theta-t_j \geq 0$, one self-consistently finds 
that the
relevant region is $\tau_j \gg \theta$. This enables one to write the average 
contribution
of a single event as:
\be
\exp\left[\left(\frac{D\theta^2 \tau}{\tau_q \tau_j}\right)^{3/2} \int d\cos 
\alpha
|\cos \alpha|^{3/2}
\int \frac{s^2 ds}{V} \left\{\frac{s^2}{{\cal G}(\frac{s}{2}\sqrt{R'})} 
\sin \frac{{\cal G}(\frac{s}{2}\sqrt{R'})}{s^2} - 1 \right\} \right]
\ee 
with now $\cos \alpha=\vec q \cdot \vec n/q$, $R'=\tau\theta^2|\cos 
\alpha|/\tau_q \tau_j^2$
and:
\be
{\cal G}(x)= \frac{2}{\sqrt{\pi}} x^3 \ e^{-x^2}.
\ee
The adequate change of variables is now found to be $\tau_j = z \theta 
\sqrt{\tau/\tau_q}$,
which means that $\tau_j$ is indeed much larger than $\theta$ for most values 
of $z$.
All the integrals (over $z, \vec n$ and $s$) can be performed exactly 
in the limit $\tau_q \ll \tau$ and the final 
result reads:
\be
S(q,\tau) = \exp\left[-\frac{2^{10} \pi^{3/8} \Gamma(-\frac{5}{4})
\Gamma(\frac{17}{8}) \sin \frac{\pi}{8}}{3^4 5^{17/8}} \ \hat \rho 
\left(\frac{\tau}{\tau_q}\right)^{5/4}\right] \qquad (\tau_q \ll \tau \ll
\theta),\label{result2}
\ee
where the ridiculously complicated numerical factor is equal to $1.00993..$. 
Therefore, we find that in this regime the $q\tau$ scaling still holds, but 
the
power $3/2$ is replaced by $5/4$. Again, the other regions of the plane 
$t',t_j$ lead
to sub-dominant corrections, the most important being $\propto \hat \rho 
(\tau/\tau_q)$.

Before discussing the `fast' collapse regime, let us insist on one further 
limitation of the
above calculation: we have assumed that the dipolar field behaves as $1/r^2$ 
everywhere in
space, that is, that the dipoles are point-like. This is valid is $q$ is large 
enough. If
$q$ is very small, the dominant source of the decay of $S(q,\tau)$ comes from 
points were
the displacement is maximum, i.e. in the immediate vicinity or `inside' the 
collapsed regions,
where the whole analysis breaks down. 

\section{The fast collapse regime} 

Let us now study the regime where $\tau \gg \theta$, corresponding to `fast' 
micro-collapses.
(Note that the fact that $\theta$ is small does not mean that the events are 
frequent:
this is described the nucleation parameter $\rho$). The calculations are very 
similar to the
above case. The relevant time scale which now appears naturally is 
\be
\tilde \tau_q= qv_0/D 
= \theta^2/\tau_q.
\ee
 When $\tau \ll \tilde \tau_q$, we find exactly the same decay as
Eq. (\ref{result2}) above: note indeed that $\theta$ actually drops out of 
this expression.
However, when $\tau \gg \tilde \tau_q$, the dynamical structure factor  
reads:
\be
S(q,\tau) = \exp\left[-\frac{16 \sqrt{2\pi^3}}{75} \rho v_0^{3/2} q^{3/2} 
\tau \right] 
\qquad (\tau \gg \tilde \tau_q),\label{result3}
\ee
which is independent of the diffusion constant $D$. Therefore, the asymptotic 
decay of $S(q,\tau)$
is a pure exponential, with an anomalous decay time $\propto q^{-3/2}$. 
This is the result one obtains if
all retardation effects are neglected (i.e. $D \to \infty$): the factor $\rho \tau$
simply counts the average number of events 
between $t$ and $t+\tau$, and $q^{3/2}$ reflects
 the fact that the distribution of local displacements $u$
  decays as $u^{-5/2}$ and has a diverging variance
   (see, e.g. \cite{Review}). 
For a distribution with a finite variance, 
one would obtain
the usual $q^2$ dependence. The appearance of power-law tails in systems with 
long-ranged
interactions is well known. It is called the Holtsmark distribution
 in the context of gravitational fields
 \cite{Chandra}.

\section{Summary of the results and discussion of the experiments}

It is useful to summarize our results in a schematic way, in terms of the 
behaviour of
$-\log S(q,\tau)$. The two physical cases depend on the
relative position of $\tau_q$ and $\tilde \tau_q$, or equivalently on the 
ratio $D\theta/qv_0$.
For $D\theta \ll qv_0$ one finds $\tau_q \ll \theta \ll \tilde \tau_q$ and:
\be 
(q\tau)^{3/2} \quad (\tau \ll \tau_q);\qquad  (q\tau)^{5/4} \quad (\tau_q 
\ll \tau \ll \tilde \tau_q);\qquad
q^{3/2}\tau \quad (\tau \gg \tilde \tau_q),
\ee
whereas for $D\theta \gg qv_0$, the $(q\tau)^{5/4}$ regime is squeezed out and 
the results are simply:
\be 
(q\tau)^{3/2} \quad (\tau \ll \theta);\qquad
q^{3/2}\tau \quad (\tau \gg \theta).
\ee
It is easy to check that the above expressions (with their corresponding 
dimensional
prefactors) correctly match one another in the crossover regions.

To which of these regimes does the experiment correspond to? First, a rather 
good
$q\tau$ scaling 
across most of the time regime ($\tau \leq 10^4$ seconds, or 3 hours). 
This means that the asymptotic regime is not observed, and therefore 
that $\max(\theta,\tilde \tau_q) > 10^4$ s. Let us first suppose that 
the experiments are in the
regime $D\theta \ll qv_0$ where $\tau_q \ll \theta \ll \tilde \tau_q$. 

According to our calculations, a $(q \tau)^{5/4}$ behaviour should be observed
for $\tau_q 
\ll \tau \ll \tilde \tau_q$. 
Fig. 1 shows that such a possibility is indeed compatible with the 
experimental data, although the data is by no means compelling. 
Under this assumption, the crossover time indicated in Fig. 1
should be interpreted 
as $q \tau_q = D \theta^2/v_0$, which is thus of order $3.\ 10^7$ cm$^{-1}$ s.
It is reasonable to 
assume that the volume of the collapsing region $v_0 \sim \xi^{3}$ is 
of the order of the 
cluster size at gelation determined in \cite{Cip}, i.e. $3. \ 10^{-8}$ cm$^3$
($\xi \simeq 30 \mu $m). 
The elastic diffusion constant can be estimated from the elastic modulus of 
the structure \cite{Cip} ($G'\simeq 10^{-3}$ dynes/cm$^2$), the value of
$\xi$ and  of the viscosity $\eta \simeq 10^{-2}$ Po:
$D=K/\gamma \simeq G' v_0/\eta \xi \simeq 10^{-6} $ cm$^2$/s. Therefore 
one gets $\theta$ on the order of
$1000$ seconds. For the maximum experimental value of $q \sim 6000$ cm$^{-1}$,
one finds $\tau_q \sim 5000$ seconds, which is a factor five larger than 
$\theta$,
in contradiction with our hypothesis. However this factor five is marginal in 
view
of the roughness of our estimates. For example, the collapsing volume $v_0$ 
could be
somewhat larger and numerical factors could help.
  
Another possibility is that the 
$(q\tau)^{5/4}$ regime is in fact not seen experimentally, i.e. 
$D\theta \gg qv_0$. This however requires that $\theta$ is actually quite 
large,
at least $10^{4}$ seconds. 
It would be interesting to determine $\rho$ or $\theta$ from an independent 
experiment in order to clarify this question.

\begin{figure}
\hspace*{+1cm}\epsfig{file=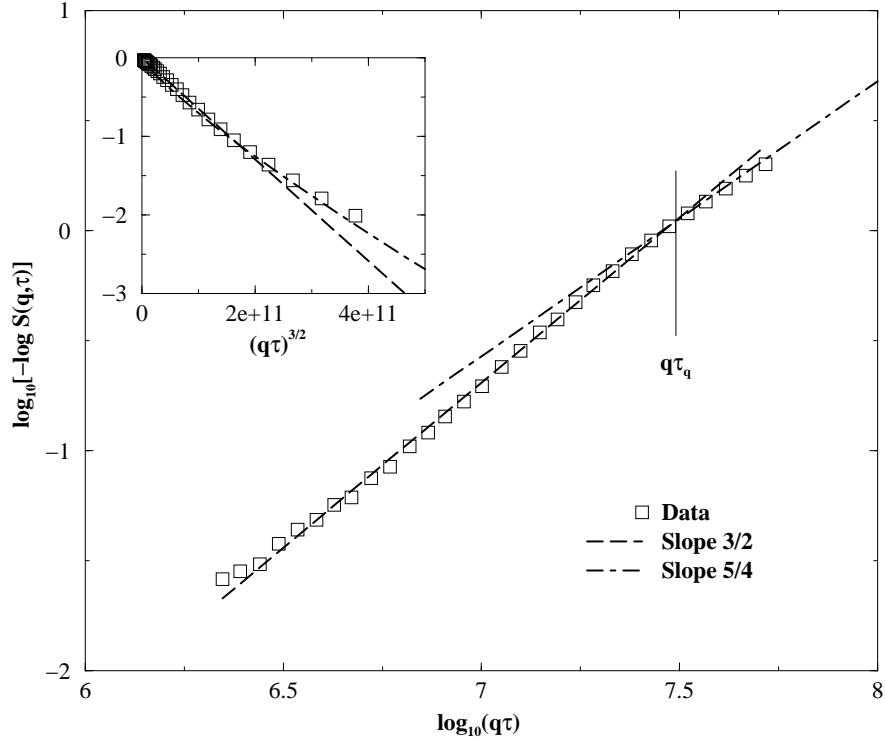,width=10cm,angle=270}
\vskip 0.3cm \caption{\small Structure factor $S(q,\tau)$ as a function of $q 
\tau$, averaged over
different $q$ (on the order of $10^{3}$ cm$^{-1}$). The data here is the same 
as that of Fig. 2,
\protect\cite{Cip}, and corresponds to a waiting time of $t_w = 67$ hours. 
Main figure: we show
$\log(-\log S(q,\tau))$ as a function of $\log q\tau$. The data is linear with 
a slope $3/2$ for most
of the time interval, but bends down at long times. This effect is more 
clearly seen in the inset,
where we have plotted $\log S(q,\tau)$ as a function of $(q \tau)^{3/2}$. A 
$(q \tau)^{5/4}$
behaviour is shown for comparison, and suggests that $q \tau_q \sim 3\ 10^{7}$ 
cm$^{-1}$ s.
\label{fig1} }
\end{figure}

\section{The aging phenomenon}

We finally discuss an important aspect of the experiments that we left out
 up to now, 
namely the fact that the dynamical structure factor is age dependent:
 the relaxation time
grows with the age $t_w$ of the sample. This growth is even unusually 
(exponentially) fast in the
initial stages \cite{Cip, Abou}.

The basic mechanism for aging is that micro-collapses induce 
a tensile strain on the non
collapsed regions. Therefore, these regions find it harder to collapse 
subsequently. 
We assume that the volume density of dipoles is $\phi(t_w)$. Therefore
$\ell^*=\phi^{-1/3}$ is the typical distance between two dipoles.
Let us
assume  also that micro-collapse events are thermally activated, 
so that their rate $\rho(t_w)$ can
be expressed as:
\be
\rho(t_w) = \rho_0 \exp\left(-\frac{\Delta E(t_w)}{k_B T}\right),
\ee
where $\Delta E(t_w)$ is an energy barrier and $T$ the temperature.
 The above mechanism means that
the barrier height depends on the  typical local strain $|\nabla u|_{typ}$.
More precisely one can write that 
$\Delta E=\frac{K}{2} (\ell_0 +|\nabla u|_{typ}\ell_0)^2
-\frac{K}{2} \ell_0^2$, where $\ell_0$ is the equilibrium length of one 
elementary
elastic unit, supposed here to be a constant.
 For small enough $|\nabla u|_{typ}$, one expects a linear 
relationship: $\Delta E \simeq K \ell_0^2 |\nabla u|_{typ}$,
 whereas for large $|\nabla u|_{typ}$ 
this dependence becomes 
quadratic: $\Delta E \simeq \frac{K}{2}\ell_0^2 |\nabla u|_{typ}^2$.
 Obviously, new events will preferentially occur in less 
strained regions, i.e., far from collapsed regions, 
inducing some spatial correlations between 
micro-collapses that we have neglected above; hence, 
we will assume in the following that the small strain regime
$\Delta E \simeq K \ell_0 |\nabla u|_{typ}$
is the dominant one. 
The typical strain  can 
be estimated as 
the sum of strains induced by all the dipoles surrounding a given
particle in the
system.
For a unique dipole of orientation $\vec n$ at a distance $r$ of the particle,
one obtains:
\be
|\nabla u|^2=\frac{v_0^2}{r^6}(3\frac{(\vec r. \vec n)^2}{r^2}+1).
\ee
From this expression, we obtain $|\nabla u|_{typ}^2=\langle |\nabla u|^2 \rangle$,
where the brackets denote the average over all contributions of dipoles
at a distance $r\geq \ell^*$ from the particle, the dipoles being
distributed with a
volume density $\phi=1/\ell^{*^3}$.
The result is
\be
|\nabla u|_{typ}=\left(\frac{8\pi}{3}\right)^{1/2} v_0 \phi
\ee
Hence, $\Delta E/k_B T=\Gamma \phi$, where $\Gamma \sim \frac{K\ell_0^2}{k_B 
T} v_0$.

We want now to evaluate the age dependence of $\rho(t_w)$ and $\phi(t_w)$.
The time evolution of $\phi$ is given by:
\be
\frac{d\phi}{dt_w} = \rho(\phi) =
 \rho_0  \exp\left(-\frac{\Delta E(\phi)}{k_B T}\right)\label{aging}
\ee
The solutions can be written explicitly and one finds that the density of
dipoles increases logarithmically with time:
\be
\phi(t_w)=\phi(0) +\frac{1}{\Gamma} 
\log \left(1+\frac{t_w}{t_0}\right),
\ee
where $t_0$ is the initial activation time $t_0=[\rho(t=0) \Gamma]^{-1}$. 
The rate of formation of dipoles (micro-collapses) therefore decreases with 
time as:
\be
\rho(t_w)=\frac{1}{\Gamma (t_w+t_0)}
\ee

The above calculations for $S(q,\tau)$ can be simply extended to 
the case where $\rho$ is not 
constant provided $\rho$ does not vary too much on the scale of $\tau$, 
which is true if $t_w$ is sufficiently large.
In order to compare with the experimentally 
determined relaxation time $\tau_r$, one must solve
the equation $S(q,\tau_r,t_w)=e^{-1}$. 
Assuming that the experiments are in the $(q\tau)^{5/4}$ regime,
one finds:
\be
\tau_r \propto (t_w+t_0)^{\mu}
\ee
with the exponent $\mu=\frac{4}{5}$. Note that the fact that
$\mu < 1$ here is a consequence of the functional form of the
non exponential relaxation. The basic mechanism, i.e. the nucleation a
micro-collapses, follows a `simple' aging law $\rho \sim t_w^{-1}$. 
Correspondingly. if the experiments are in the 
$(q\tau)^{3/2}$ regime or in the final predicted regime in
$q^{3/2} \tau$, $\mu$ would respectively have the values
$\mu=\frac{2}{3}$ and $\mu=1$. The best agreement with the experimentally 
determined $\tau_r \sim t_w^{0.9}$ \cite{Cip} is for $\mu=\frac{4}{5}$.

Thus, as observed experimentally, $\tau_r$ seems to saturate to a finite value 
$t_0^\mu$ for $t_w \to 0$ and to grow as $t_w^{4/5}$ for long times (see 
\cite{Cip}, Fig. 5). All this results have been obtained in the regime of small strains.
If the strains become large (i.e. at large times), the energy becomes 
quadratic in
$|\nabla u|_{typ}$ and reads $\Delta E/k_B T=\Gamma' \phi^2$,
where $\Gamma' \sim \frac{K\ell_0^2}{k_B T}v_0^2$.
In this case, logarithmic corrections appear in the previous expressions
and one obtains in the large time regime: $\tau_r \propto t_w^{\mu} 
\log^{\mu/2}[\rho_0 \sqrt{\Gamma'} t_w]$. More generally, if $\Delta E \propto \phi^\nu$,
one finds an initial regime where $\tau_r$ grows as $\exp(t_w^\nu)$
at first, and then as $t_w^\mu$ with logarithmic corrections.

Finally, the above mechanism suggests that when the density of dipoles 
becomes of 
order unity,
the system should macroscopically collapse, as indeed seen experimentally
\cite{Cip} at long times.

\section{Conclusion}

In conclusion, we have computed the dynamical structure factor $S(q,\tau)$ 
of an elastic medium where
force dipoles appear at random in space and time, as `micro-collapses'
 appear in the structure. The balance between these collapse events and the 
elastic
 relaxation of the  internal stresses in the medium lead to
various regimes, depending on the
wave vector $q$ and the collapse time $\theta$. 
In an early time regime, the logarithm
of the structure factor behaves as $(q\tau)^{3/2}$ plus subleading 
corrections, as
anticipated in \cite{Cip} using heuristic arguments.
 However, in an intermediate time that 
might be relevant for the experiments of \cite{Cip},
 we obtain a $(q \tau)^{5/4}$ behaviour 
that would have been difficult
to guess from simple arguments. Finally, the asymptotic long time regime 
is found to behave
as $q^{3/2} \tau$, where the $q\tau$ scaling is not obeyed; but this last 
regime seems
not to be observed experimentally and it is likely that before that, a 
macroscopic
collapse occurs as the one observed in colloidal gels \cite{Cip}.

The relevance of the microscopic mecanism described in this paper still 
remains to be
checked experimentally, in particular in order 
to know if the micro-collapses, if they indeed exist, continuously occur in 
time and space or
if they only occur in a short time interval $t_c$ after the formation of the 
gel.
We have made a calculation similar to the one above for this last case. 
In the limit where $\theta \ll t_w$, we find $-\log S(q,t_w,t_w+\tau) 
\propto
t_c (q\tau/t_w^{5/4})^2$ for $\tau \ll t_w$ which then saturates to a finite 
value
(which goes to $1$ as $t_w \to \infty$, since in this limit nothing moves any longer).
 
We have also given a plausible scenario for aging,
in terms of a strain dependent energy barrier for micro-collapses. 
The relaxation time is found to grow with the age $t_w$,
exponentially at first, and then as $t_w^{4/5}$
with logarithmic corrections, which seems to be in good agreement with 
experiments.

It would be interesting to analyze other experiments where
similar effects have been reported \cite{Lairez, Lequeux, Cipelse}, in 
particular in
micellar crystalline phases and onion phases,
along the lines of the present work. In these systems however, 
motion of defects such as dislocations or grain boundaries may also play a 
crucial role.

\vskip 1cm 

{\it Acknowledgements}$\ \ $ 
J.P. B. wants to thank D. S. Fisher and Harvard University for hospitality 
during the period this work was performed. 
We are indebted to L. Cipelletti, S. Manley, L. Ramos and 
D. Weitz for very important comments and for providing some experimental data.
We also thank B. Abou, M. Adam, C. Caroli, J.P. Carton, D. Lairez, F. Lequeux 
and G. Porte for discussions on related matters.

\end{document}